\documentclass[twocolumn]{aastex701}

\usepackage{url}
\usepackage{xcolor}
\newcommand{\Msol}[1]{M\textsubscript{\(\odot\)}}

\begin{document}

\title{Kinematic Signatures in the Stellar Halo\\ from Cosmological Encounters between the Milky Way and its Clouds}

\author[0009-0007-8066-0832]{Mia Mansfield}
\affiliation{Department of Physics \& Astronomy, University of Pennsylvania, Philadelphia, PA 19104, USA}
\affiliation{Institute for Astronomy, University of Hawai`i, 2680 Woodlawn Drive, Honolulu, HI 96822, USA}
\email{mansm@sas.upenn.edu}

\author[0000-0003-3939-3297]{Robyn Sanderson}
\affiliation{Department of Physics \& Astronomy, University of Pennsylvania, Philadelphia, PA 19104, USA}
\email{robynes@sas.upenn.edu}

\author[0000-0003-3244-5357]{Daniel Hey}
\affiliation{Institute for Astronomy, University of Hawai`i, 2680 Woodlawn Drive, Honolulu, HI 96822, USA}
\email{danielhey@outlook.com}

\author[0000-0001-8832-4488]{Daniel Huber}
\affiliation{Institute for Astronomy, University of Hawai`i, 2680 Woodlawn Drive, Honolulu, HI 96822, USA}
\email{huberd@hawaii.edu}

\author[0000-0002-8354-7356]{Arpit Arora}
\affiliation{Department of Astronomy, University of Washington, Seattle, WA 98195, USA}
\email{arora125@sas.upenn.edu}

\author[0000-0002-6993-0826]{Emily Cunningham}
\affiliation{Department of Astronomy, Boston University, 725 Commonwealth Ave., Boston, MA 02215, USA}
\email{eccunnin@bu.edu}

\author[0000-0001-5214-8822]{Nondh Panithanpaisal}
\affiliation{Carnegie Observatories, 813 Santa Barbara St, Pasadena, CA 91101, USA}
\affiliation{TAPIR, California Institute of Technology, Pasadena, CA 91125, USA}
\email{npanithanpaisal@carnegiescience.edu}

\defcitealias{Cunningham_2020}{C20}
\begin{abstract}
Recent theoretical and observational analysis of the interaction between the Milky Way (MW) and LMC suggest that it has a significant dynamical impact on the MW's stellar halo. We investigate this effect using simulations from the \textit{Latte} project, a simulation suite from the Feedback In Realistic Environments 2 (FIRE-2) Project. By comparing simulations with and without an LMC-analog interaction, we show that fully cosmological LMC interactions create prominent velocity asymmetry in the stellar halo of the MW, resulting from both barycentric displacement (the "reflex motion") and the dynamical wake of the LMC. The strength and direction of this asymmetry depend on the mass ratio at pericenter and orbit of the LMC analog. We perform a spherical-harmonic decomposition of the velocities of halo star particles to confirm that the identified signatures are LMC-induced and persist even when LMC star particles are removed. We also show that this strategy separates and individually detects the dipole ($\ell=1$) of the global reflex motion and the quadrupole ($\ell=2$) of the local wake. These asymmetries are consistent with those identified in previous work using non-cosmological simulations; the dipole is easily distinguishable from other complex halo substructure using spherical harmonics while the quadrupole is sometimes confused. These findings support the detectability of MW--LMC interaction signatures in upcoming observational surveys of the MW stellar halo.
\end{abstract}

\keywords{\uat{Milky Way stellar halo}{1060} --- \uat{Large Magellanic Cloud}{903} --- \uat{Milky Way dynamics}{1051}}


\section{Introduction} \label{sec:intro}

The LMC and SMC are among the Milky Way's (MW) most massive and well known satellite galaxies. The two form a binary pair, with the LMC estimated to be roughly 10$\times$ more massive than the SMC, accounting for ~10-20\% of the MW's total mass \citep{Garavito2021, Vasiliev2023_LMC, chandra2024all}. In addition to their large combined mass, key observations such as a lack of a stellar tidal stream, numerous satellite galaxies, and ongoing star formation suggest that the L/SMC are on their first orbital infall into the MW \citep{besla2007magellanic,besla2010simulations, Vasiliev2023_LMC}. 

Recent advances in our understanding of the MW's Clouds come alongside \textit{Gaia} DR3 data suggesting the MW is in dynamical disequilibrium, exhibiting large-scale kinematic asymmetries \citep{besla2007magellanic, besla2010simulations, hunt2025milky, gaiaDr3}. Given its large mass and close approach, the LMC is expected to have contributed significantly to the observed MW perturbations \citep{vasiliev2021tango, Vasiliev2023_LMC, hunt2025milky}. The most prominent signature of LMC influence is a global effect that induces MW barycenter shifting. This causes a reflex motion throughout the MW's stellar halo: while the inner halo can quickly react by moving towards the LMC, stars in the outer halo lag behind due to their longer orbital time scales \citep{Gomez2015, GaravitoCamargo2019HuntingForDMWake,Cunningham_2020, Garavito2021, Arora2023}. The resulting asymmetry impacts the stellar kinematics, dark matter (DM) distribution, and overall alignment of the MW halo \citep{Petersen_2020, Garavito2021, Shipp2021_streamMass, Baptista_2022,Arora2023, chandra2024all}. In addition to global effects, the LMC has a local effect on the MW known as the dynamical friction wake. This leads to an over-density of stars and DM trailing the LMC's path and influences MW tidal streams in the vicinity \citep{Garavito2021, Shipp2021_streamMass,  brooks2024lmccalls}. This influence is studied in depth for the Sagittarius stream in \citet{vasiliev2021tango}.

Observational data from surveys such as \textit{Gaia} and SDSS have allowed significant progress to be made towards constraining the LMC's influence \citep{Petersen_2020, conroy2021all, chandra2024all}. However, existing surveys of the stellar halo lack the observational density needed to isolate these effects \citep{Auge_2020, Hey_2023}. 

Stellar variability has long been recognized as a powerful distance tracer for stars in the LMC \citep{Tabur2010}. Recent advances in all-sky ground such as ATLAS \citep{Tonry2018_ATLAS} and ASAS-SN \citep{Kochanek2017_ASAS-SN} now enable such distance measurements to probe dynamics of stars in the galactic bulge \citep{Hey_2023} and dwarf satellites \citep{ToguchiTani2025_SagWithGaia}. \citet{Auge_2020} estimates that roughly 70,000 M Giant candidates for asteroseismology exist in the stellar halo, extending to radial distances of up to 200 kpc. Additionally, models of Large Synoptic Survey Telescope (LSST) surveys predict the detection of a few thousand RR Lyrae (RRL) stars at radial distances greater than 100 kpc \citep{Sanderson_2017_NewViews}. Observations of both M Giant and RRL stars at these distances would capture previously unconstrained signatures of LMC infall. 

In the absence of sufficient observational data, past works constraining the effects of a massive satellite have relied on simulations of the MW-LMC system. Works such as \citet{Cunningham_2020}, \citet{Garavito2021}, and \citet{brooks2024lmccalls} have used constrained simulations to isolate the MW halo response to LMC infall. While analysis of simplified MW-LMC models has extensively increased our understanding of the large scale impacts of LMC infall, predictions from these simulations cannot be directly extrapolated to observational data. This motivates the need for predictions from realistic cosmological-baryonic simulations which can resolve different stellar populations. Simulations such as those developed in the Feedback In Realistic Environments-2 (FIRE-2) Project\footnote{\url{http://fire.northwestern.edu}}, present a way to predict the observable signatures of massive satellite infall ahead of observational data \citep{Wetzel2016_simulatingSatellites, Hopkins2014_FIRE}. 

We examine the predictability of MW response signatures using simulations from the \textit{Latte }project – a suite of FIRE-2 simulations. The \textit{Latte} project simulates MW-like systems at ultra high resolutions, and therefore can resolve halo substructure and simulate realistic populations of satellite galaxies \citep{Wetzel2016_simulatingSatellites}. Simulations from the \textit{Latte} suite are uniquely suited for understanding the formation and evolution of MW-mass galaxies, making them the ideal tool for examining MW-LMC interaction.


\section{Data and Methods} \label{sec:data}

\subsection{Latte Simulations} \label{subsec:sims}
A comprehensive list of simulation and infall properties can be found in Table \ref{tab.infall_props}. 

We use five halos, namely \textbf{m12b}, \textbf{m12c}, \textbf{m12f}, \textbf{m12w}, and \textbf{m12i}, from the \textit{Latte} suite of zoomed-in cosmological-baryonic simulations of MW-mass ($~10^{12} \Msol{}$) galaxies, part of the Feedback In Realistic Environments (FIRE-2) project \citep{Wetzel2016_simulatingSatellites, Wetzel_2023_Data_Release}.  \footnote{These simulations are publicly available \citep{Wetzel_2023_Data_Release} at \url{http://flathub.flatironinstitute.org/fire}.} These simulations are run using the GIZMO code \citep{Hopkins2015_mesh} \footnote{\url{http://www.tapir.caltech.edu/~phopkins/Site/GIZMO.html}} and employ the the FIRE-2 physics model \citep{Hopkins2018_FIRE2}.  A full description the FIRE-2 physics model and numerical methods used to solve gravity can be found in \citet{Hopkins2018_FIRE2} and \citet{Wetzel_2023_Data_Release} for the specific implementation.

FIRE-2 galaxies have been shown to resemble the MW in stellar mass, gas content, DM mass, and density profiles \citep{Wetzel2016_simulatingSatellites, Hopkins2018_FIRE2, Sanderson_2020_SyntheticGaiaSurveys}. The selected halos have total masses of approximately $1$--$1.5 \times 10^{12} \Msol{}$ and are resolved with initial particle masses of $m_\mathrm{b} = 7100 \ \Msol{}$ for stars and gas, and $m_\mathrm{DM} = 35,000 \ M_{\odot}$ for DM in the zoomed-in region. Each star particle is representative of a small population of stars with the same ages and abundance. Star particles follow the population through typical stellar evolution.

Of the selected simulations, \textbf{m12b} is most analogous to the observed MW--LMC system \citep{Arora2023, garavito2024corotation, Arora2025Shaping} in its orbital properties and satellite-to-mass ratio (1:8). The LMC analog in \textbf{m12b} has a first pericentric distance of $r_\mathrm{peri} \sim40$ kpc and an infall mass of $ = 1.2 \times 10^{11} M_\odot$, comparable to the estimated values of $r_\mathrm{peri} = 45 \pm0.5$ kpc and $m_{peri} = 1-2.5 \times 10^{11} M_\odot $ for the LMC \citep{Vasiliev2023_LMC}.

Although \textbf{m12b} has a large satellite interaction that is comparable to the MW--LMC system, the LMC analog has a much earlier pericenter time of T=8.81 Gyr \citep{Arora2023}. In addition to the known differences between the simulated and physical LMCs, there are many features of the LMC that remain uncertain, such as its orbit, proper motion, and center of mass \citep{Vasiliev2023_LMC}. To account for potential discrepancies in the signatures found in our simulations and future observational data, we consider three more FIRE-2 halos: \textbf{m12c}, \textbf{m12f}, and \textbf{m12w}. These simulations are chosen for the presence of massive satellites with mergers comparable to the LMC in terms of pericenter mass, infall time, and infall radius. These simulations, along with \textbf{m12b}, are identified by \citet{arora2022Stability, Barry_2023} as containing an LMC analog. In simulation \textbf{m12f}, the large satellite has a similar orbital trajectory to the physical LMC, but is only half as massive \citep{Arora2023}. Simulations \textbf{m12w} and \textbf{m12c} also contain massive satellites that are similar to the LMC. However, these satellites have very different orbits than the LMC, leading to infall signatures that may differ from those expected \citep{Arora2023}. The last selected simulation, \textbf{m12i}, lacks a massive satellite galaxy and acts as a control \citep{Hopkins2018_FIRE2, Baptista_2022}. 

We define the principal axis of each simulation using a subset of central, young stars. We first select the 25\% youngest stars within 10 kpc of the galactic center, then refine the selection to retain only the 90\% most centrally concentrated stars by mass. Using centralized young stars ensures that the principal axis accurately reflects the host galaxy's present day morphology.

\subsection{Removing LMC-Analog Contamination} \label{subsec:pres_day_snapshots}

In order to detect the LMC's impact on the MW, we first identify and remove the LMC analog in each simulation. Removal of the LMC analog is essential to isolating signatures of its influence. The MW-analog has stellar components that extend to 100 kpc from the center of the galaxy, and the LMC analog is less than 40 kpc away from the galactic center at present day. Therefore, it is difficult to discern evidence of the LMC analog's influence from the LMC analog itself. In order to combat this problem, we create a pipeline to identify and remove star particles belonging to the LMC analog in each simulation. 

In order to mimic the physical MW--LMC system, we define present-day as 50 Myr after the first pericentric passage. We select present-day relative to pericenter time rather than the LMC analog's position because of the strong impact that pericenter time has on the MW halo response, which differs significantly before and after LMC infall \citep{Arora2025Shaping}. Moreover, the dipole signature corresponding to halo reflex motion peaks roughly 50 Myr after pericenter \citetext{Darragh-Ford, in prep}. Therefore, selecting present-day with respect to pericenter time allows us to accurately capture the expected signal strength in the MW-analog halo. 

For simulation \textbf{m12i}, we define present-day to be at the same snapshot used for simulation \textbf{m12b}. This allows us to directly compare simulations that have evolved for the same length of time, rather than following simulation \textbf{m12i} to z=0. We take present-day from simulation \textbf{m12b} because it is most analogous to the physical MW-LMC system.

In simulations \textbf{m12b}, \textbf{m12w}, \textbf{m12c}, and \textbf{m12f}, we track the LMC analog until present-day using merger trees and halo catalogs. We adapt the framework in \citet{Panithanpaisal_2021} to isolate a single satellite of choice. From there, we follow the LMC analog through all snapshots where it can be isolated. We begin tracking at a time after the LMC analog has absorbed the majority of infalling satellites, but before it forms the majority of stars at present day. We then iterate through all snapshots between the start of tracking and pericenter. At each snapshot, we identify the LMC analog and add any unique star indices within it to a list.

Despite our rigorous tracking of the LMC analog, we find that the iterative selection method alone does not sufficiently isolate the massive satellite. Specifically, this selection misses stars from many of the satellite galaxies that infall into the LMC analog before tracking begins. To alleviate this issue, we identify every halo that merges with the LMC analog. We then use the same pipeline to track each halos from formation to infall, adding these stars to the catalog.

While tracking merging satellite galaxies captures most stars, we still found that the isolated LMC analog was missing 15-20\% of of its total stars. In order to remedy this discrepancy, we consider all star particles within 15 kpc of the LMC analog's location at any given point to be part of the massive satellite, regardless of whether or not they are recognized by the satellite tracking pipeline. This method inevitably erroneously identifies some stars from other satellites as part of the LMC analog, and also leaves a gap at the present day location. However, the complete removal of the LMC analog from the host galaxy is essential to our analysis, while the absence of a few accreted halo stars and a relatively small cap near the infall region will not significantly impact any dynamic signatures. 

Our final selection of LMC-analog stars includes those obtained through both methods of halo tracking and nearby star selection.

\begin{figure*}[htbp]
    \centering
    \includegraphics[width=1\textwidth]{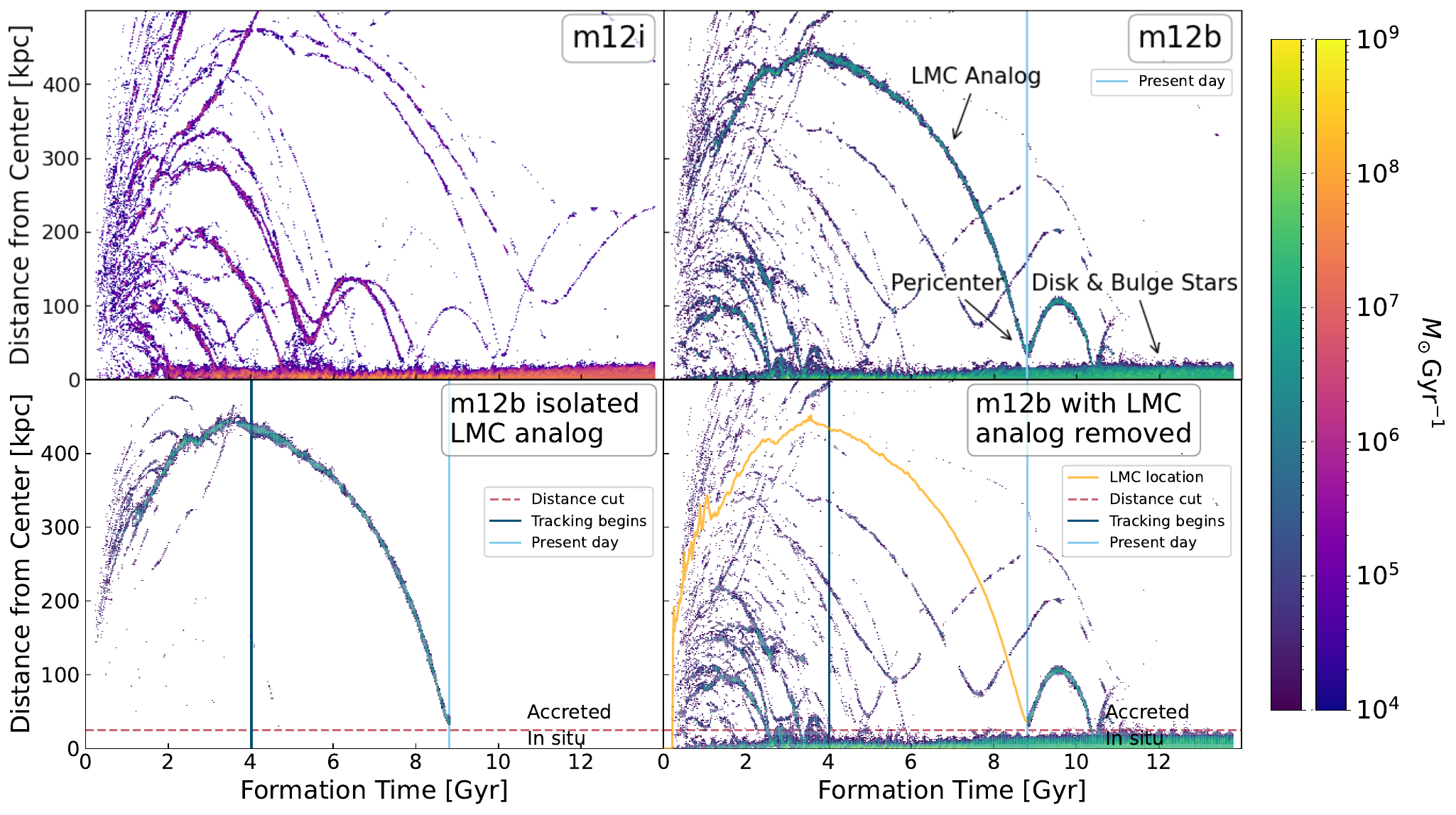}
    \caption{Formation distance vs time of formation for star particles in simulation \textbf{m12i} (top left), simulation \textbf{m12b} (top right), the isolated \textbf{m12b} LMC analog (bottom left), and \textbf{m12b} without the LMC analog (bottom right).  Individual satellite galaxies appear as thin lines, and the host galaxy's disk appears as a dense bar below 25 kpc. We define 8.81 Gyr as present day because of the comparable location and trajectory of the LMC analog and current LMC. In all analysis of simulations with an LMC analog, we use the selection with the LMC analog removed.}
    \label{fig:formation_dist}
\end{figure*}

We verify the accuracy of this separation by examining the star formation history of the host galaxy and LMC analog. As an example, we show the star formation distance vs time for simulations \textbf{m12b} and \textbf{m12i} (Figure \ref{fig:formation_dist}). Star forming satellite galaxies can be identified in Figure \ref{fig:formation_dist} as thin lines, with the thickness corresponding to the relative size of the satellite. The dense bar between 0 kpc and 25 kpc represents all stars formed within the host galaxy's disk and bulge. In the formation history of \textbf{m12b} (Figure \ref{fig:formation_dist}, top right), the LMC analog appears as a much thicker line than other satellite galaxies because of the large number of stars that form within it. Its relative distance from the MW can be traced as stars form along its path. 

The isolated LMC analog and host galaxy selections for simulation \textbf{m12b} can be seen in the bottom row of Figure \ref{fig:formation_dist}. The LMC-analog selection initially included trace disk stars from the MW analog. We remove these star particles by considering all stars formed less than 25 kpc from the center of the host galaxy to be part of the disk, and therefore cut them out of the LMC-analog selection in analysis. This distance limit is indicated as the dashed line in the bottom row of Figure \ref{fig:formation_dist}. The resulting lists for simulation \textbf{m12b} contain 9,409,523 host galaxy star particles and 1,460,441 LMC-analog star particles.

\begin{figure*}[thbp]
    \centering
    \includegraphics[width=1\textwidth]{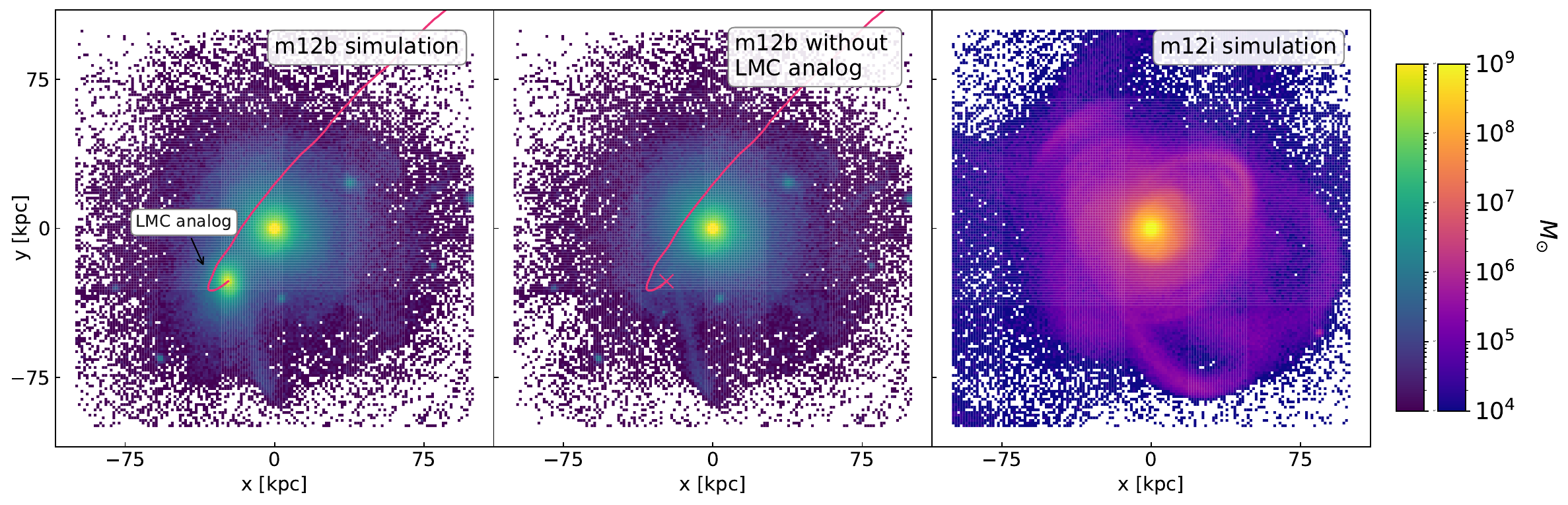}
    \caption{
    Mass weighted distribution of stars particles in \textbf{m12b} with (left) and without (center) the LMC analog and in \textbf{m12i} (right) simulated galaxies. Galaxies are face on in the xy plane. In \textbf{m12b} before removal, the LMC analog can be identified as the smaller galaxy $\sim40$ kpc from the center. The LMC analog's infall trajectory is shown in all \textbf{m12b} plots as an orange line.} 
    \label{fig:visual}
\end{figure*}

Our selections leave a thorough removal of the LMC analog. The results of the removal can be seen for simulation \textbf{m12b} in the center plot of Figure \ref{fig:visual}, and for all other simulations in the right-most column of Figure \ref{fig:velocity_cartesian}.

In addition to the LMC analogs present in simulations \textbf{m12b}, \textbf{m12c}, \textbf{m12f}, and \textbf{m12w}, we note that our control, \textbf{m12i}, has a stellar stream extending through the host galaxy resulting from a merger that differs from LMC infall (Figure \ref{fig:visual} right). A stream of this size will have significant dynamic effects on its host galaxy, but we expect these perturbations to differ from those induced by a massive satellite. Therefore, we use \textbf{m12i} as a control to compare the influence of a large satellite against typical galactic evolution.

\begin{table*}[htbp]
    \centering
   
    \begin{tabular}{llllllll}
        \hline
        \hline
        Simulation   & 
        $R_{*, 90}$ [kpc] & $M_{*, 90}$ [$M_{\odot}$] & $z_{\mathrm{peri}}$ & $t_{\mathrm{peri}}$ [Gyr] & $r_{\mathrm{peri}}$ [kpc] & $m_{\mathrm{peri, LMC}}$ [$M_{\odot}$]& $m_{\mathrm{peri, MW}}$ [$M_{\odot}$]\\
        \hline
        \textbf{m12b} & 3.62 & 6.20E+10 & 0.49 & 8.81 & 37.90 & 1.2E+11 & 3.81E+11\\
        \textbf{m12c} & 9.36 & 1.16E+11 & 0.07 & 12.89 & 18.13 & 5.08E+10 & 2.12E+11\\
        \textbf{m12f} & 11.11 & 1.43E+11 & 0.26 & 10.79 & 35.74 & 6.02E+10 & 4.08E+11\\
        \textbf{m12w} & 10.79 & 1.00E+11 & 0.59 & 7.97 & 7.67 & 4.89E+10 & 6.01E+10\\
        \textbf{m12i} & 9.10 & 8.63E+10  &      &      &      &          &  \\
        \hline
        \hline
    \end{tabular}
     \caption{Properties of MW-mass galaxies with massive satellite interactions in FIRE-2 simulations. Properties are defined at the time of LMC-analog pericenter as denoted by $t_{\mathrm{peri}}$.}
    \parbox{\linewidth}{\raggedright
    $R_{*, 90}$ Radius of the host galaxy enclosed within a spherical radius that encompasses 90\% of the stellar mass within 20kpc 

    $M_{*, 90}$: Total mass enclosed within radius $R_{*, 90}$ 
    
    $z_{\mathrm{peri}}$, $t_{\mathrm{peri}}$, $r_{\mathrm{peri}}$: Time and distance from the host galaxy's center of the LMC analog's peri-centric passage

    $m_{\mathrm{peri, LMC}}$: LMC-analog total mass at time of peri-centric passage

    $m_{\mathrm{peri, MW}}$: Host galaxy total mass enclosed within radius $r_{peri}$.
    }
    \label{tab.infall_props}
\end{table*}

\subsection{Spherical Harmonics} \label{subsec:sph_eqs}

LMC induced effects are large scale perturbations, unlike the small scale perturbations caused by substructure. In order to isolate these effects, we follow the methodology in \citet{Cunningham_2020} (hereafter \citetalias{Cunningham_2020}) to apply spherical harmonic expansions to velocity signatures on within the host galaxy's stellar halo. This allows us to quantify kinematic variations on different spatial scales and identify infall signatures.  We use the spherical harmonic expansion notation from \citetalias{Cunningham_2020} as follows.

Laplace's spherical harmonics with order $\ell$ and degree $m$ are defined as:
\begin{equation}
    Y_\ell^m (\theta, \phi) = \sqrt{\frac{2\ell + 1}{4\pi}\frac{(\ell - m)!}{(\ell + m)!}}P_\ell^m (cos\theta)e^{im\phi},
\end{equation}
with colatitude $\theta$ measured from North to South, and azimuthal angle $\phi$ measured along the x-y plane. The associated Legendre polynomials are $P_\ell^m$.

The spherical harmonics define a complete set of orthogonal functions. Therefore, they form an orthonormal basis for any function $f(\theta,\phi)$ on the surface of a sphere:

\begin{equation}
    f(\theta, \phi) = \sum_{\ell=0}^{\infty}\sum_{m=-\ell}^{\ell} a_{\ell m}Y_\ell^m(\theta, \phi),
\end{equation}
where the $a_{\ell m}$ spherical harmonic coefficients are defined by:

\begin{equation}
    a_{\ell m} = \int_{\Omega} f(\theta, \phi) Y_\ell ^{m*} (\theta, \phi) \mathrm{d}\Omega.
\end{equation}

We focus on the $a_{\ell m}$ spherical harmonic coefficients and the $C_\ell$ angular power spectrum throughout our analysis. These coefficients define the dominant $\ell$ and $m$ modes in each signal, reflecting which events have the largest impact on the MW analog's velocity profile.

Using the spherical harmonic coefficients $a_{\ell m}$, the angular power spectrum $C_\ell$ can be computed as:
\begin{equation}
    C_\ell = \frac {1}{2\ell + 1}\sum_m |a_{\ell m}|^2.
\end{equation}

Each value of $\ell$ has $2 \ell + 1$ associated $m$ values. Therefore, the total power of a given $\ell$ is $(2\ell + 1) \times C_\ell$. We use this definition of power, rather than $C_\ell$, for power spectra throughout our analysis.

Following \citetalias{Cunningham_2020}, we rely on the \texttt{healpy} \citep{healpy} Python package for all spherical harmonic analysis. All spherical harmonic coefficients and power spectra are computed using the \texttt{healpy} function \texttt{anafast}. We use the \texttt{healpy} mapping function \texttt{mollview} to visualize the spherical harmonics and velocity signatures as Mollweide projections. 

\section{Results} \label{sec:results}

\subsection{Kinematic Maps} \label{subsec:cartesian_kin}

\begin{figure*}[htbp]
    \centering
    \includegraphics[width=0.9\textwidth]{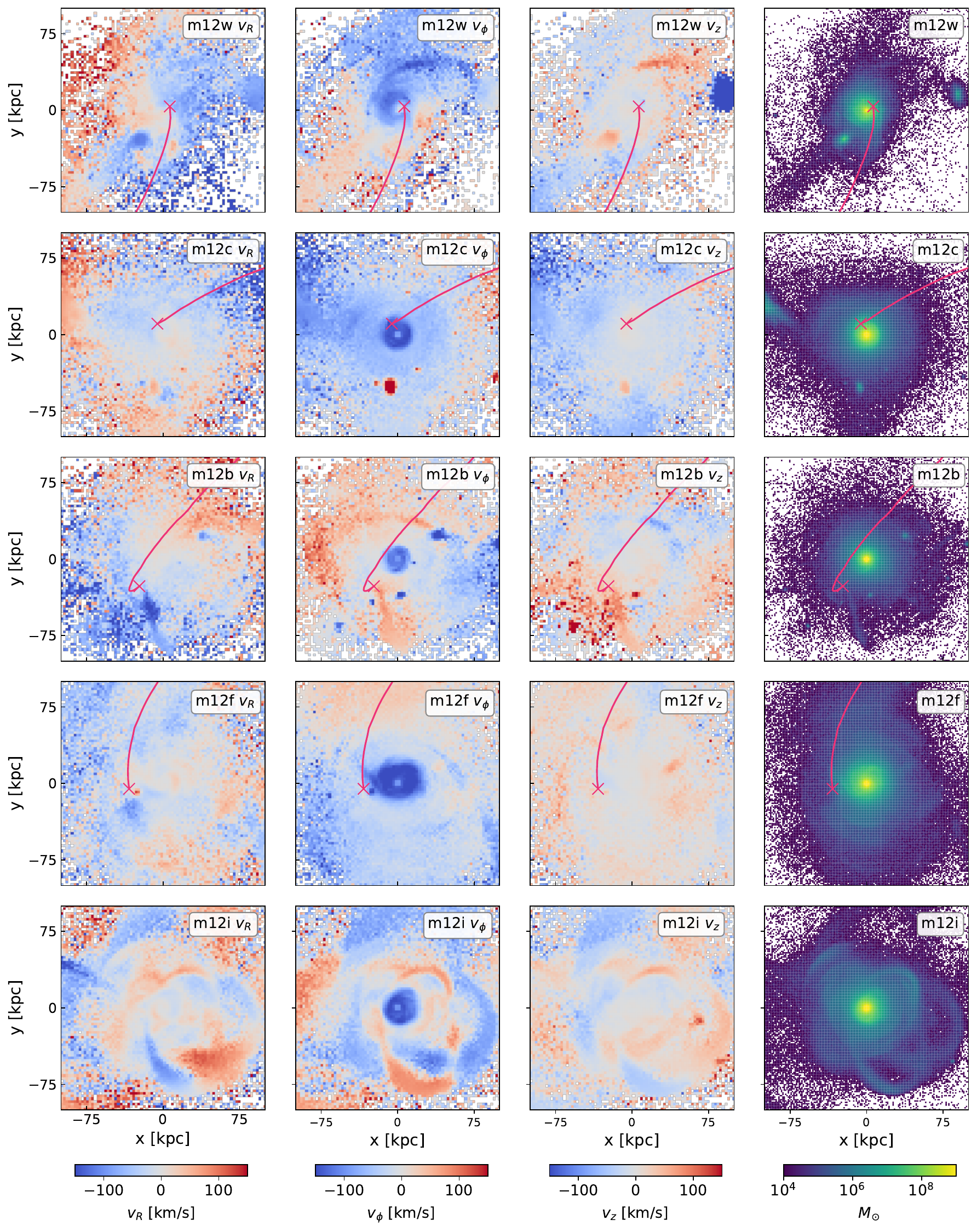}
    \caption{Face-on views of cylindrical velocities for the stellar halo of host galaxies of \textbf{m12w}, \textbf{m12c}, \textbf{m12b}, \textbf{m12f}, and \textbf{m12i} simulations. In simulations with an LMC analog, the most notable infall signature is a global kinematic asymmetry. This asymmetry is most prominent in the radial velocity signature (left column). We include all stars within a vertical slab of 100kpc above and below the disk plane to capture features of the stellar halo. Star particles belonging to the LMC analog are removed in our analysis, and the LMC analog's trajectory is displayed as a pink line. } 
    \label{fig:velocity_cartesian}
\end{figure*}

In order to evaluate the kinematic effects of the LMC's infall, we examine cylindrical velocity distributions of host galaxy stars for all simulations. The LMC analog is removed in all analysis. The resulting velocity maps can be seen in Figure \ref{fig:velocity_cartesian}.

While velocity distributions from isolated, face-on spiral galaxy are expected to be relatively smooth, tidal streams, satellite galaxies, and other substructure can alter a galaxy's dynamics. The size and morphology of each kinematic irregularity alludes to the dynamic interaction that caused it, allowing us to probe the host galaxy's evolutionary history, and, in this case, identify the influence of an LMC-sized satellite.

The most prominent perturbation caused by the LMC analog is a global asymmetry in radial velocity (Figure \ref{fig:velocity_cartesian}, left column). This signature results from reflex motion, as has been well established in previous works (e.g. \citet{Arora2023}). This pattern is identifiable in $v_R$ maps for all simulations with an LMC analog. While the asymmetry is slightly visually obstructed by substructure in all simulations, it is especially obscured for simulation \textbf{m12f} due to stellar streams within the halo. Notably, \textbf{m12i} does not display the same global asymmetry due to the lack of a large satellite. Instead, it features several long, narrow regions of coherent radial velocity, which result from the presence of a large stellar stream.

In the $v_\phi$ component (Figure \ref{fig:velocity_cartesian} second column), \textbf{m12w}, \textbf{m12f}, and \textbf{m12b} exhibit an asymmetrical pattern similar to the radial velocity asymmetry. However, unlike the other simulations with LMC analog, simulation \textbf{m12c} displays a relatively uniform azimuthal distribution. This suggests that the reflex motion asymmetry in \textbf{m12c} is weaker than in other simulations, amounting to only a radial asymmetry and lacking a north-south component. This is because the satellite's orbit is almost exclusively within the disk plane, which results in an infall that does not induce a strong reflex motion. The specifics of this orbit and absence of a strong asymmetry are discussed further in \citet{Arora2023}.

In contrast to the simulations that have LMC analogs, \textbf{m2i} does not display either global asymmetry or uniformity. As with the radial component, regions of azimuthal perturbation are driven by the large stellar stream.

Across all simulations, kinematic patterns are the least prominent in $v_z$ maps (Figure \ref{fig:velocity_cartesian}, right column). This is expected, as stars primarily orbit within the disk plane and therefore are less susceptible to vertical perturbations. While local responses to infall signatures are often present in vertical perturbations, these signatures are more difficult to qualitatively identify compared to global signatures. The only simulation with an LMC analog and a notable vertical velocity signature is \textbf{m12b}, which contains a local perturbation near the satellite infall region. This is due to the \textbf{m12b} LMC analog having a near perpendicular infall trajectory relative to the desk, resulting in a slight vertical kinematic response. Simulation \textbf{m12i} also contains regions of slight vertical perturbation as a result of the stream, though these signatures are significantly weaker in magnitude than the $v_R$ and $v_\phi$ perturbations.

Although many kinematic features are identifiable in Figure \ref{fig:velocity_cartesian}, local effects from the LMC analog are not always observable, and global signatures are still difficult to unambiguously interpret by inspection alone. The presence of stellar streams, dwarf galaxies, and other substructure complicate velocity fields, making it challenging to isolate specific contributions from the LMC analog. Therefore, we apply a spherical harmonic expansion to the velocity fields to confirm that the identified signatures are responses to the LMC analog's influence.

\subsection {Spherical Harmonic Expansion} \label{subsec:sph_expans}

\begin{figure*}[htbp]
    \centering
    \includegraphics[width=0.9\textwidth]{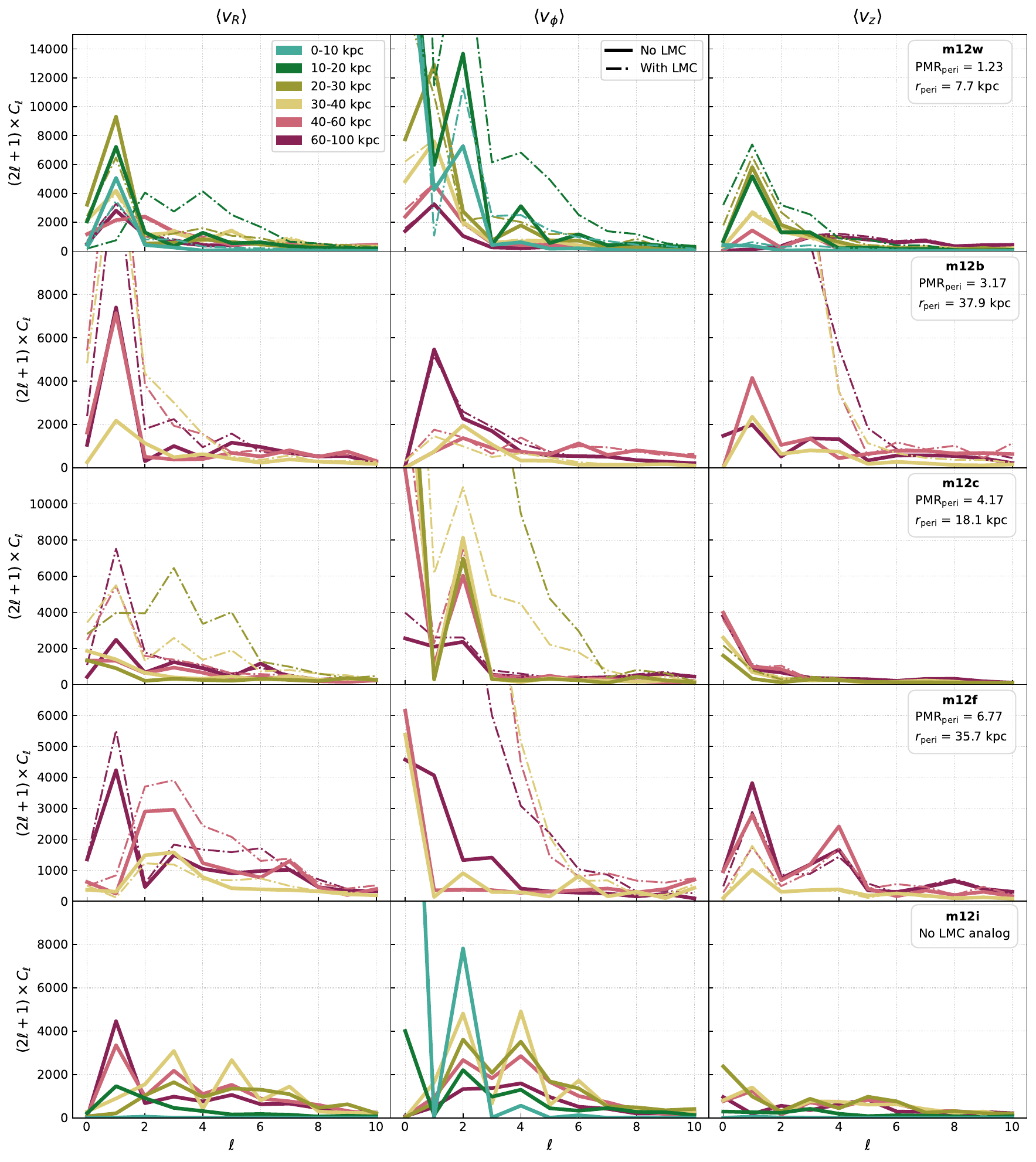}
    \caption{Power spectra for mean cylindrical velocity signatures in simulations \textbf{m12b}, \textbf{m12w}, \textbf{m12f}, \textbf{m12c}, and \textbf{m12i} (rows 1--5 respectively). In all simulations with an LMC analog, the global reflex motion is evident as a high $\ell = 1$ power in at least the radial velocity signature (left column), though for most simulations the strong power persists through the other signatures as well. Dashed lines reflect the power spectra with the LMC analog included, whereas solid lines have the LMC analog removed. Spectra are computed for spherical shells at increasing radii, starting from the infall region and extending outward in 10 kpc increments to 40 kpc, followed by 40--60 kpc and 60--100 kpc shells. We highlight the differing $y$-axis ranges for each simulation, which reflect differing powers based on the simulation's pericenter mass ratio ($\mathrm{PMR}_{\mathrm{peri}}$) and radius ($r_{\mathrm{peri}})$} 
    \label{fig:Cl_power}
\end{figure*}

\begin{figure*}[htbp]
    \centering
    \includegraphics[width=0.9\textwidth]{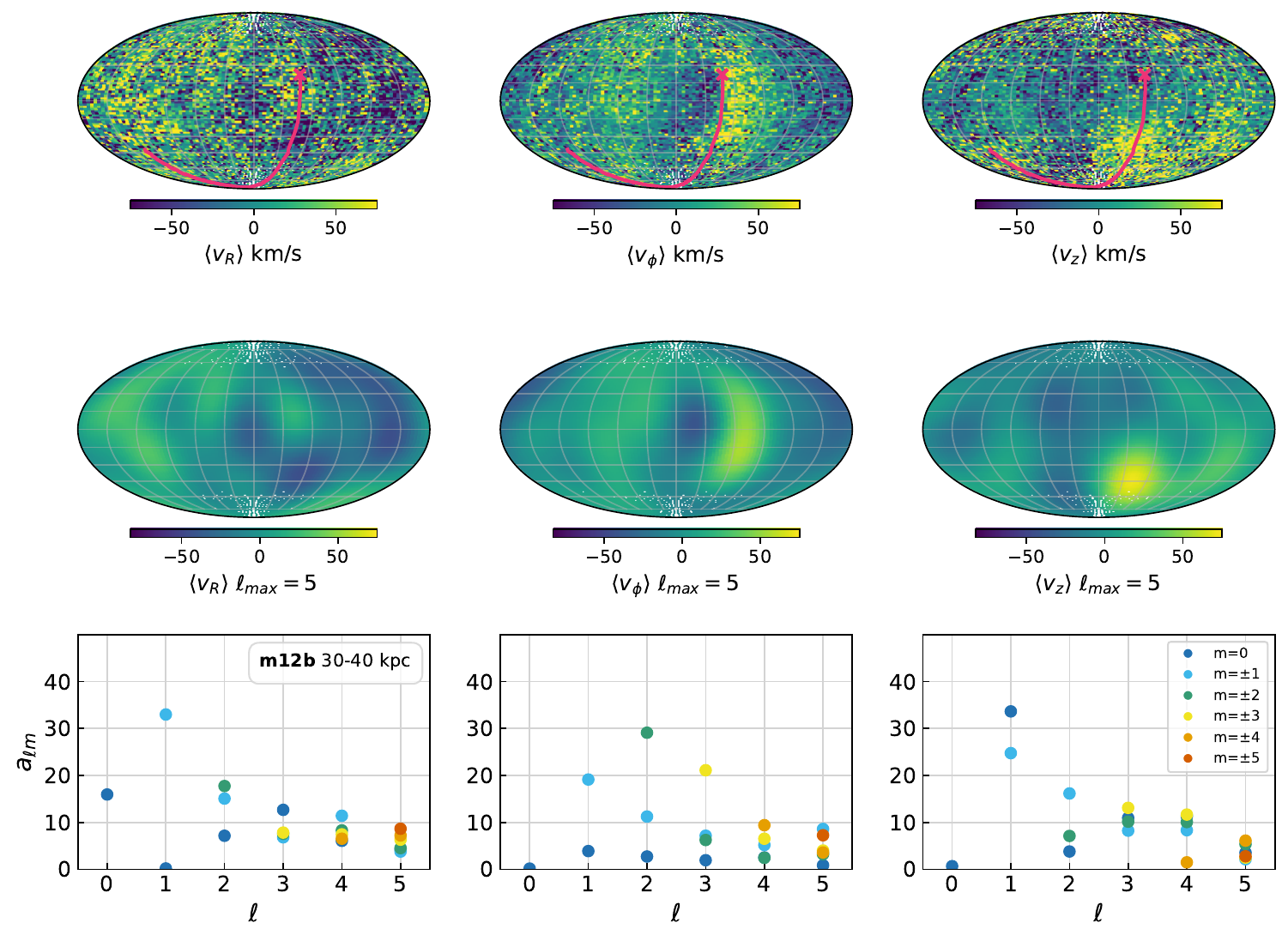}
    \caption{Top row: Average velocity maps for cylindrical velocity $v_R$ (left), $v_\phi$ (center), and $v_z$ (right) of \textbf{m12b} stars located 30--40 kpc from the galactic center. The angular trajectory of the LMC analog is marked in red, with its present-day location marked by an X.
    \parbox{\linewidth}\raggedright Center row: Spherical harmonic expansion for the average velocity maps above. The maximum expansion term is set at $\ell_{\mathrm{max}} = 5$ to highlight low order features.
    Bottom row: Magnitude of $a_{\ell m}$ coefficients in the spherical harmonic expansion. Different colors correspond to degrees of $m$. In both radial and vertical velocity, the dominant mode is $\ell=1$, capturing the dynamic asymmetries that arise due to the global reflex motion. The azimuthal velocity is dominated by the $\ell = 2$, $m= \pm 2$ mode, which captures a local, outward motion resulting from the local dynamical friction wake.} 
    \label{fig:alm_and_sph_expansion}
\end{figure*}

We confirm that kinematic patterns in the host galaxy can be attributed to the LMC analog through spherical harmonic expansions. LMC-analog induced perturbations are captured by the lowest spherical harmonic modes: the $\ell=0$ mode at small radii reflects bulk disk shifting, $\ell=1$ mode captures dipole asymmetry from global reflex motion, and the $\ell=2$ mode reflects local perturbations caused by the dynamical friction wake. A more detailed description of these modes is given in \citetalias{Cunningham_2020}. We examine the angular power spectrum of each simulation at a variety of radii in order to identify the effects of large satellite infall at varied radial distances. We then perform a closer examination of the spherical harmonic expansion near the pericenter region to separate local and global infall signatures.

\subsubsection{Global Velocity Features}

We analyze the power spectra of mean cylindrical velocity signatures to isolate the dynamic effects of an LMC analog (Figure \ref{fig:Cl_power}). Each simulation is divided into concentric spherical slices, beginning at a radius encompassing the infall region with 10 kpc widths, or at the galactic center for \textbf{m12i}. We increase the width of each spherical shell at larger radii to account for decreased stellar density. The LMC analogs are removed from all simulations. Each simulation is analyzed on a different power range, corresponding to the $y$-axes in rows of Figure \ref{fig:Cl_power}. This allows us to account for differing power amplitudes across simulations, since we are primarily concerned about relative power between $\ell$ modes. The amplitude differences between simulations are informed by the LMC analog's pericentric stellar mass ratio and infall distance: a closer pericentric passage and smaller pericenter mass ratio results in stronger perturbations.

The $v_R$ power spectrum in simulations \textbf{m12w}, \textbf{m12b}, and \textbf{m12f} reflects the identified global asymmetry through a prominent $\ell=1$ mode (Figure \ref{fig:Cl_power}, left column). In most simulations, this mode peaks at large radii, consistent with results from \citetalias{Cunningham_2020}. However, because simulation \textbf{m12w} has short pericenter radius, the interior asymmetries have not yet fully propagated to the outer halo, resulting in slightly lower powers in the $\ell=1$ mode at large radii. Simulation \textbf{m12c} does not have strong $\ell=1$ mode, reflecting the lack of a strong global asymmetry. Although \textbf{m12c} does contain an LMC analog, this result is unsurprising due to the satellite's disk-aligned infall trajectory.

The $v_\phi$ signature in simulations \textbf{m12w}, \textbf{m12c}, and \textbf{m12f} exhibits a strong $\ell=0$  mode. This mode is most prominent at inner radii, capturing the bulk rotational motion of the disk and inner halo. The absence of this mode in simulation \textbf{m12b} can be attributed to the disk in \textbf{m12b}, which is more condensed and therefore does not have a rotational signature that extends to the radii we are examining. In addition to the $\ell=0$ mode, all simulations with LMC analogs have peaked $\ell=1$ modes at large radii. This captures the same reflex motion asymmetry as is seen in $v_R$.

All simulations with an LMC analog also exhibit an azimuthal $\ell=2$ mode at radii near infall. This is reflective of local perturbations caused by the LMC analog. The power of the $\ell=2$ is notably less than that of $\ell=0$ and $\ell=1$, indicating that the LMC's local effect is less prominent than global effects. As such, local kinematic features are not easily identifiable by inspection alone, as can be seen by the lack of prominent perturbations near infall regions in Figure \ref{fig:velocity_cartesian}. Thus, we conclude that local perturbations alone cannot confirm the presence of an LMC analog.

The vertical velocity distribution in simulations \textbf{m12w}, \textbf{m12b}, and \textbf{m12f} also exhibits a strong $\ell=1$ mode, reflecting the dipole asymmetry seen in radial velocity. Across all simulations, low-order $v_z$ modes exhibit lower powers than in radial and azimuthal signatures. Unlike \textbf{m12w}, \textbf{m12b}, and \textbf{m12f}, simulation \textbf{m12c} does not have the same strong $\ell=1$ mode in $v_Z$ due to the weaker reflex motion effect. We note that unlike the results of \citetalias{Cunningham_2020}, we don't see a significant $\ell=0$ monopole term in vertical or azimuthal velocity as a result of global shifting. However, it is likely that substructure within the halo obscures this potential signal. Additionally, we can confidently identify a global effect from the high power $\ell=1$ terms alone, so the lack of a detectable monopole in all signatures is not a significant concern.

In addition to the low-order modes, many simulations contain power in higher-order modes, which reflect perturbations caused by other halo substructure. For example, we note the strong $\ell=4$ mode in azimuthal velocity for \textbf{m12w}, as well as the strong $\ell=4$ mode in the vertical velocity of \textbf{m12f}. These peaks can be attributed to the dynamical influence of smaller satellite galaxies near the host halo. 

We compare the angular power spectrum of simulation \textbf{m12i} (Figure \ref{fig:Cl_power}, bottom row) to the spectra from simulations with an LMC analog. As expected based on the results of \citetalias{Cunningham_2020}, the stellar stream in \textbf{m12i} causes high powers across alternating modes, creating a jagged power spectrum corresponding to kinematic asymmetries on all scales. This power spectrum differs significantly from those caused from LMC analogs, confirming that the identified low-order signals and lack of a persistent pattern through all powers is a result of the LMC analog's influence. However, simulation \textbf{m12i} does contain a high power in the $v_\phi$ $\ell=2$ mode. This quadrupole term is also seen in simulations with an LMC analog. Therefore, because this term appears in the power spectrum of simulations with and without an LMC analog, an $\ell=2$ mode is not necessarily indicative of a dynamical friction wake and cannot be used as an indicator of massive satellite infall.

\subsubsection{Velocity Features near Infall} \label{subsubsec:vel_near_infall}
We next examine the velocity field near the region of the LMC analog through a spherical harmonic expansion. Velocity signatures in this region are expected to highlight local effects from large satellite infall.  As an example, we discuss these features for simulation \textbf{m12b}. Spherical harmonic expansions for the other simulations are explored in the Appendix. Mean cylindrical velocity maps, spherical harmonic expansions, and values of spherical harmonic coefficients are displayed for simulation \textbf{m12b} in Figure \ref{fig:alm_and_sph_expansion}.

A mollweide projection of the mean velocity distribution of simulation \textbf{m12b} is pictured in cylindrical components in the top row of Figure \ref{fig:alm_and_sph_expansion}. The angular trajectory and position of the LMC analog is indicated in red, where patterns in $v_{\phi}$ local to the path show evidence of the local response. The center row of Figure \ref{fig:alm_and_sph_expansion} shows a spherical harmonic expansion of the velocity maps with $\ell_{\mathrm{max}}=5$. This expansion captures the dominant signatures in each velocity component, reflecting large scale patterns seen in the velocity maps above.

We present the magnitude of each spherical harmonic coefficient in the bottom row of Figure \ref{fig:alm_and_sph_expansion}. The radial velocity signature (left column) is dominated by the $\ell=1, m=\pm1$ mode, which captures the global shifting caused by the LMC analog. As is discussed for the angular power spectrum of $v_R$, this signature is stronger at larger radii. However, the asymmetry is still present at inner radii. The $v_{\phi}$ velocity maps (center column) are dominated by a quadrupole $\ell=2, m=\pm2$ signature, which reflects the dynamical friction wake in the LMC analog infall region. Similar to $v_R$, the $v_z$ velocity signature (right column) has a dominant $\ell=1, m=0$ mode. This reflects a global north-south asymmetry in vertical velocity, where the northern and southern halo components both shift towards the LMC analog respectively.


\section{Conclusions} \label{sec:conclusion}

In this work, we use five FIRE-2 hydrodynamical zoom-in simulations to ensure that predicted signatures of an LMC-like satellite are observable in a realistic galactic environment. We identify and compare kinematic perturbations in 4 LMC-analog infall scenarios. Furthermore, we use a spherical harmonic expansion to characterize kinematic signatures in the stellar halo. Our main findings are as follows:

\begin{itemize}
    \item The most prominent signature of large satellite infall is a radial velocity asymmetry, resulting from reflex motion. While this signature can often be identified by inspection, a low order spherical harmonic expansion confidently captures the dipole $\ell=1$ mode in agreement with \citetalias{Cunningham_2020}.
    
    \item The power of infall signatures is largely dependent on the relative size and position of the LMC analog at pericenter. A small stellar mass ratio and close infall radius contribute to a stronger reflex motion effect, resulting in a more dramatic dipole asymmetry. 
    
    \item Local velocity perturbations from the dynamical friction wake are less distinct than the global reflex motion. While all of the simulations with an LMC analog exhibit power in the $\ell=2$ mode, corresponding to local perturbations, the power of these modes is significantly less than the power of $\ell=1$ modes. Additionally, perturbations from other halo substructure can lead to comparable power in the $\ell=2$ mode, as is seen in simulation \textbf{m12i}. Thus, while quadropole wake ($\ell=2$) perturbations are detectable in all simulations, power in the $\ell=2$ mode alone is not a reliable signature for identifying MW-LMC interaction.
    
\end{itemize}

While our findings confirm the prevalence of previously predicted signatures of LMC-infall in realistic galactic environments, the observability of these findings in various stellar tracers is still uncertain. 

As future surveys of the MW's outer disk and stellar halo extend to larger radii, the detection of signatures of LMC influence will become increasingly feasible. Current and upcoming asteroseismic data of M Giants may allow us to catalog stars up to 200 kpc away \citep{Auge_2020, Hey_2023}, reaching well into regions of LMC influence. The relative abundance and luminosity of M Giant populations provide a promising solution to current observational restraints, opening opportunities to identify predicted effects.

Continued investigation of cosmological simulations can can prepare us for the forthcoming stellar catalogs. In particular, examining the observability of both global and local signatures in various stellar tracers can provide more accurate predictions ahead of observational results. Tools such as \texttt{py-ananke} \citep{Thob_2024} can generate mock stellar catalogs and further constrain the observable signatures of LMC infall. Thus, continuing to examine the effects of large satellite infall will better prepare us for the wealth of observational data to come.

\section*{Acknowledgements}
\label{sec:Acknowledgements}

The authors thank Alex Riley and Nico Garavito-Camargo for helpful discussion and comments.

MM acknowledges support from Research Experience for Undergraduate program at the Institute for Astronomy, University of Hawai'i-Mānoa funded through NSF grant \#2050710. MM would like to thank the Institute for Astronomy for their hospitality during the development of this project. The development of this project was made possible by the computing cluster resources provided by the Flatiron Institute Center for Computational Astrophysics.

MM, RES, D. Hey, D. Huber, and AA are grateful for support from NSF grant AST-2007232. MM and RES were also supported by Sloan Foundation award FG-2023-20669. RES additionally acknowledges support from Simons Foundation grant 1018462 that contributed to the development of this project.
D.H. acknowledges support from the National Aeronautics and Space Administration (80NSSC22K0781) and NASA’S Interdisciplinary Consortia for Astrobiology Research (NNH19ZDA001N-ICAR) under award number 19-ICAR19 2-0041. 
A.A. acknowledges support from Gordon and Betty Moore Foundation, the DiRAC Institute in the Department of Astronomy and the eScience Institute, both at the University of Washington.

\software{This project made use of GizmoAnalysis \citep{GizmoAnalysis} and HaloAnalysis \citep{HaloAnalysis} which were first used in \citet{Wetzel2016_simulatingSatellites}, Healpy \citep{healpy}, Numpy \citep{numpy}, Scipy \citep{scipy}, and Matplotlib \citep{matplotlib}.}


\newpage
\bibliographystyle{aasjournal}


\newpage
\section*{Appendix}
\label{sec:Appendix}
In this appendix, we repeat the analysis in Section \ref{subsubsec:vel_near_infall} to examine the dominant kinematic features near infall for simulations \textbf{m12w}, \textbf{m12c}, \textbf{m12f}.

\begin{figure*}[htbp]
    \centering
    \includegraphics[width=1\textwidth]{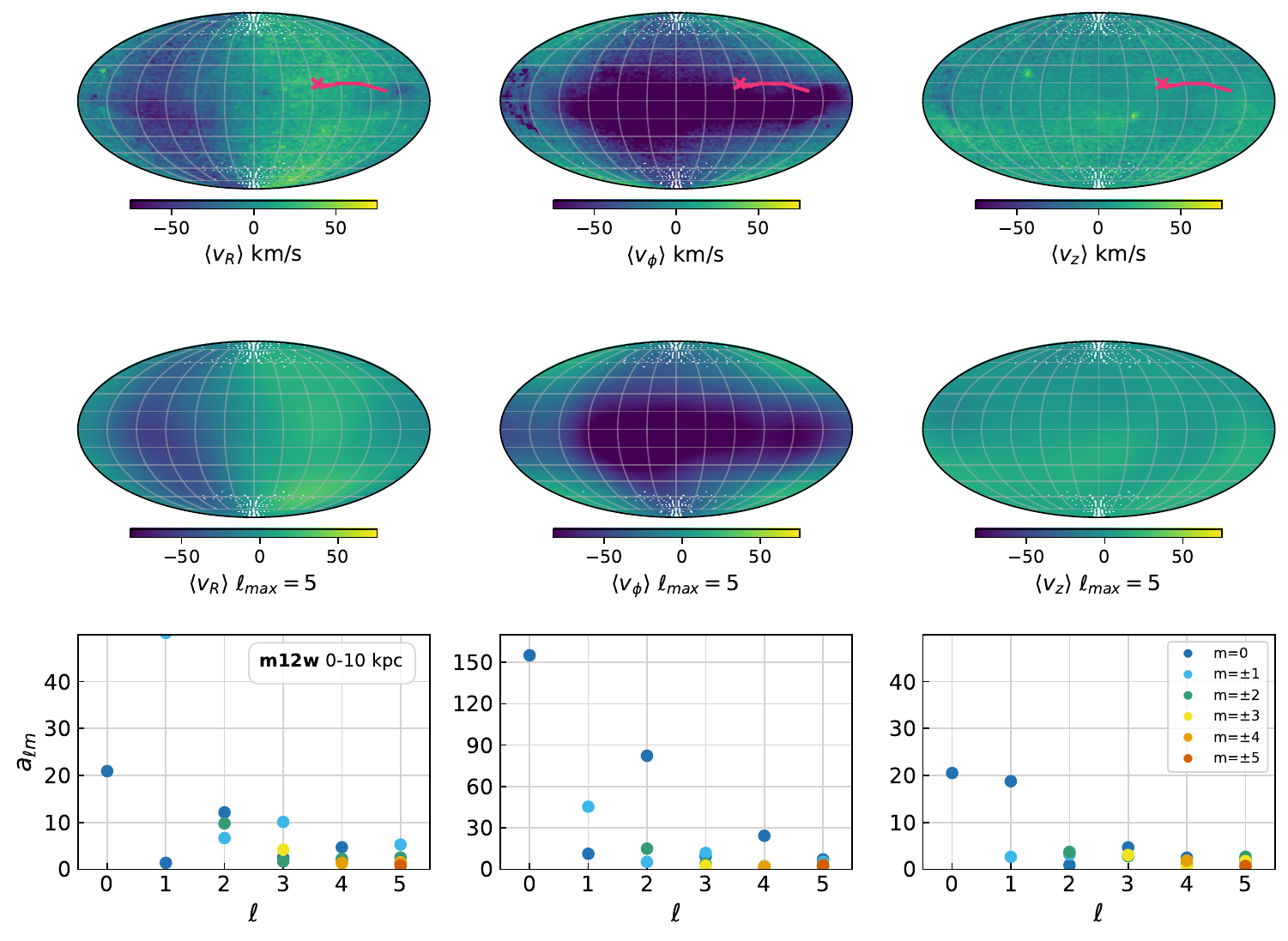}
    \caption{Top row: Average velocity maps for cylindrical velocity $v_R$ (left), $v_\phi$ (center), and $v_z$ (right) of \textbf{m12w} stars located 0-10 kpc from the galactic center. The angular trajectory of the LMC analog is marked in red, with its present-day location marked by an X.
    \parbox{\linewidth}\raggedright Center row: Spherical harmonic expansion for the average velocity maps above. The maximum expansion term is set at $\ell_{\mathrm{max}} = 5$ to highlight low order features.
    Bottom row: Magnitude of $a_{\ell m}$ coefficients in the spherical harmonic expansion. Different colors correspond to degrees of $m$. The $v_R$ component is dominated by the $\ell=1, m=1$ mode, reflecting the bulk radial shifting of the disk towards the LMC analog. In $v_\phi$, the $\ell=0, m=0$ dominates, capturing the high rotational power of the disk. The $v_z$ component is also dominated by the $\ell=0, m=0$ mode, which results from bulk upward shifting of the disk towards the LMC analog's location.} 
    \label{fig:m12w_alm_and_sph_expansion}
\end{figure*}

Mollweide projections of cylindrical velocity signatures, a spherical harmonic expansion, and values of the spherical harmonic coefficients for simulation \textbf{m12w} are pictured in Figure \ref{fig:m12w_alm_and_sph_expansion}. The LMC analog in this simulation has a pericenter radius of $<8$kpc, and we therefore examine the innermost 10kpc of the host galaxy. This results in fairly uniform radial velocity maps and corresponding spherical harmonic expansions. In the $v_R$ component (left column), the asymmetry caused by reflex motion is very apparent. This is further confirmed by the high power $\ell=1, m=\pm1$ mode, which captures a radial velocity dipole. The $v_\phi$ signal in \textbf{m12w} is largely uniform, as it captures the rotation of the disk. This corresponds to the monopole $\ell=0, m=0$ mode. The same mode dominates the $v_z$ signature, reflecting a bulk upward motion of the disk towards the LMC analog.

\begin{figure*}[htbp]
    \centering
    \includegraphics[width=1\textwidth]{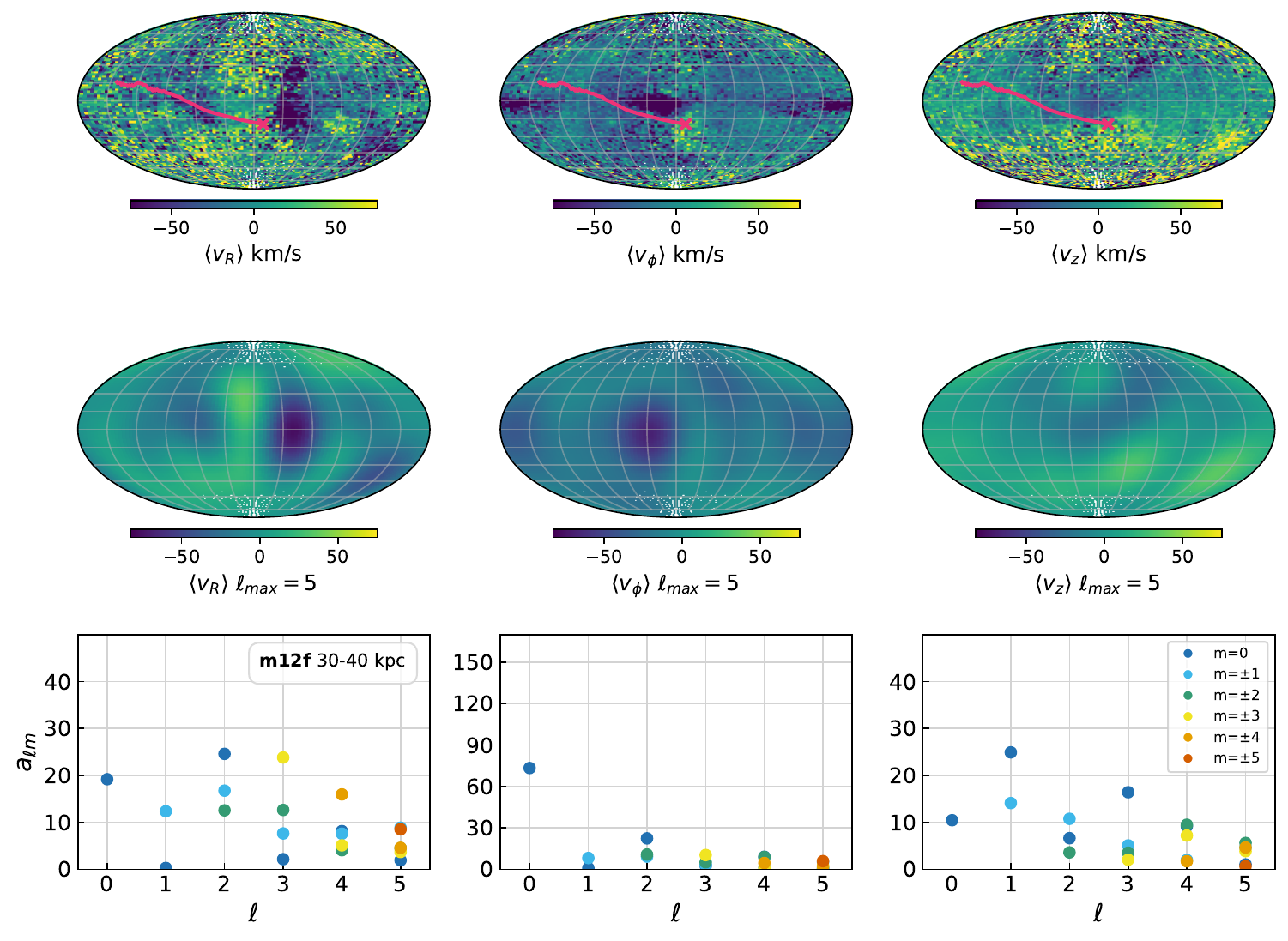}
    \caption{Top row: Average velocity maps for cylindrical velocity $v_R$ (left), $v_\phi$ (center), and $v_z$ (right) of \textbf{m12f} stars located 30-40 kpc from the galactic center. The angular trajectory of the LMC analog is marked in red, with its present-day location marked by an X.
    \parbox{\linewidth}\raggedright Center row: Spherical harmonic expansion for the average velocity maps above. The maximum expansion term is set at $\ell_{\mathrm{max}} = 5$ to highlight low order features.
    Bottom row: Magnitude of $a_{\ell m}$ coefficients in the spherical harmonic expansion. Different colors correspond to degrees of $m$. The $v_R$ component is dominated by the $\ell=2, m=0$ mode, which captures the outward motion of stars responding to the local dynamical friction wake. The dominant $v_\phi$ mode is in $\ell=0, m=0$, reflecting bulk rotation at this radius. The $v_z$ component is dominated by the $\ell=1, m=0$ mode, which reflects the global asymmetry caused by reflex motion.} 
    \label{fig:m12f_alm_and_sph_expansion}
\end{figure*}

The same cylindrical velocity maps and expansions are pictured in Figure \ref{fig:m12f_alm_and_sph_expansion} for simulation \textbf{m12f}. Because \textbf{m12f} has a larger pericenter radius than simulation \textbf{m12w}, there is significantly more substructure present in the radial slice of interest. This results in more disrupted velocity maps. The most obscured signal is $v_R$ (left column). This signature still contains power in the $\ell=1, m=\pm1$ mode corresponding to a reflex motion signature, but is primarily dominated by lower power modes. The dominant $\ell=2, m=0$ and $\ell=3, m=\pm3$ modes are caused by substructure within the halo, rather than being evidence of a large satellite. The $v_\phi$ signature also lacks strong signals from the LMC analog, and is instead dominated by the $\ell=0, m=0$ mode associated with disk rotation. However, the global asymmetry is present in $v_z$, and is captured by the dominant $\ell=1, m=\pm1$ mode.

\begin{figure*}[htbp]
    \centering
    \includegraphics[width=1\textwidth]{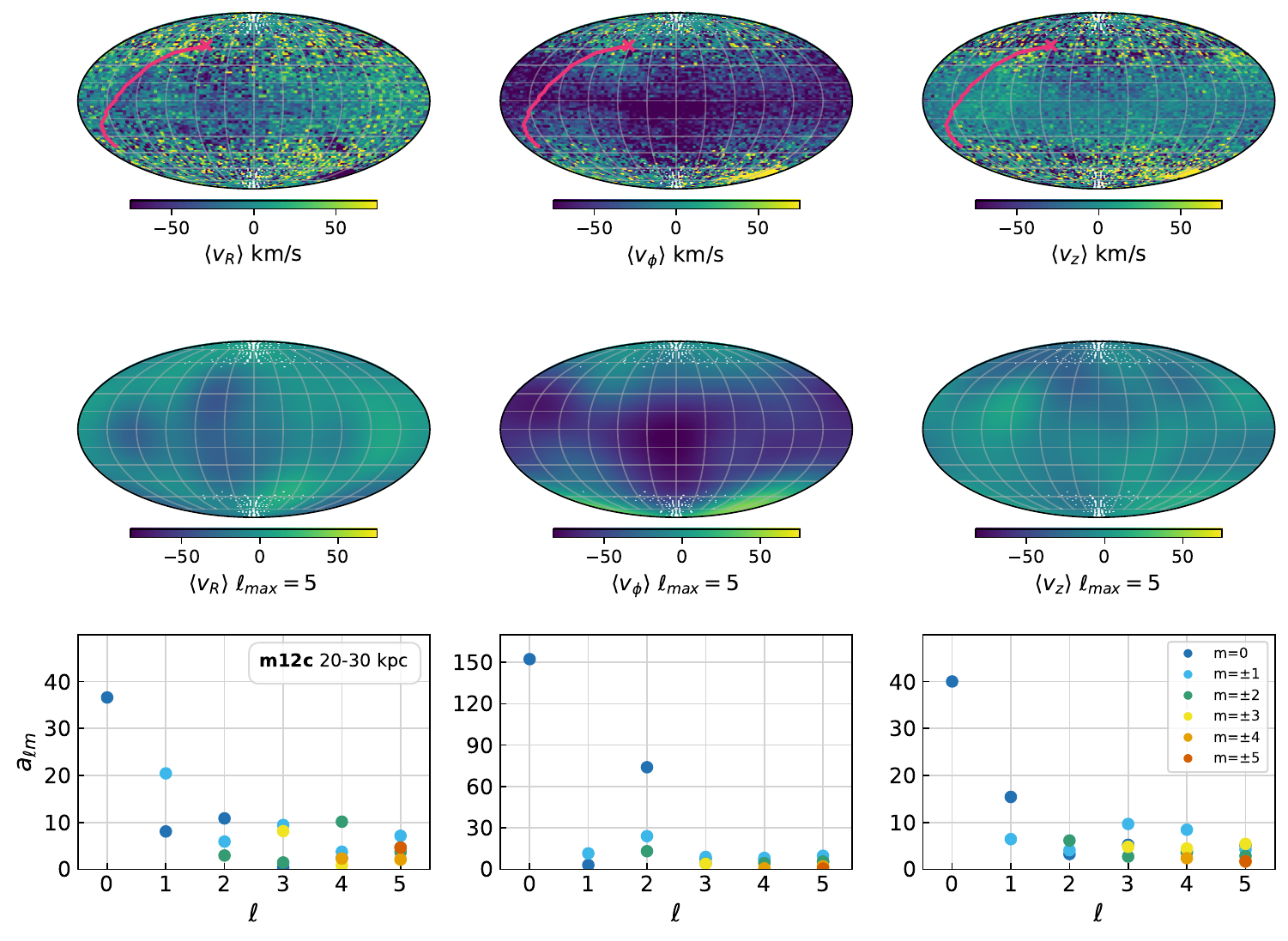}
    \caption{Top row: Average velocity maps for cylindrical velocity $v_R$ (left), $v_\phi$ (center), and $v_z$ (right) of \textbf{m12c} stars located 20-30 kpc from the galactic center. The angular trajectory of the LMC analog is marked in red, with its present-day location marked by an X.
    \parbox{\linewidth}\raggedright Center row: Spherical harmonic expansion for the average velocity maps above. The maximum expansion term is set at $\ell_{\mathrm{max}} = 5$ to highlight low order features.
    Bottom row: Magnitude of $a_{\ell m}$ coefficients in the spherical harmonic expansion. Different colors correspond to degrees of $m$. All velocity components are dominated by the $\ell=0, m=0$ mode, reflecting bulk motion throughout the galaxy. Due to the LMC analog's infall trajectory and therefore lack of reflex motion, \textbf{m12c} has a much weaker $\ell=1, m=\pm1$ radial mode than the other simulations.} 
    \label{fig:m12c_alm_and_sph_expansion}
\end{figure*}

Lastly, we discuss the spherical harmonic expansion for simulation \textbf{m12c} (Figure \ref{fig:m12c_alm_and_sph_expansion}). The \textbf{m12c} LMC-analog trajectory is primarily within the disk plane, and therefore the satellite does not induce many significant perturbations. This can be seen in the mean velocity distributions (top row) and spherical harmonic expansions (Center row), which are generally uniform. This is further confirmed by the spherical harmonic coefficients. All velocity components in \textbf{m12c} are dominated by the $\ell=0, m=0$ mode, corresponding to a monopole signal. We note that the power of this mode is significantly higher for $v_\phi$, reflecting the rotational motion of the outer disk. We also note that unlike the other simulations with LMC analogs, \textbf{m12c} does not have a strong $\ell=1, m=\pm1$ mode that would correspond to the global reflex motion. This aligns with the results of the \textbf{m12c} angular power spectrum (Figure \ref{fig:Cl_power} third row), which contains a low power $\ell=1$ mode.

\end{document}